\documentclass{article}
\usepackage{spconf,amsmath,epsfig}

\usepackage{amssymb}
\usepackage{bbm}
\begin{document}\sloppy

% Example definitions.
% --------------------
\def\x{{\mathbf x}}
\def\L{{\cal L}}

\newcommand{\remove}[1]{}

% Title.
% ------
\title{Attention and Visibility in an Information-Rich World}
%
% Single address.
% ---------------
\twoauthors
{Nathan O. Hodas}
{University of Southern California\\ 
Information Sciences Institute \\ 
Marina Del Rey, CA\\
nhodas@isi.edu}
{Kristina Lerman\thanks{This material is based upon work supported in part by the
Air Force OfÞce of ScientiÞc Research under Contract Nos.
FA9550-10-1-0569, by the National Science Foundation under Grant No. CIF-1217605, and by DARPA under Contract
No. W911NF-12-1-0034.}}
{University of Southern California\\ 
Information Sciences Institute \\ 
Marina Del Rey, CA\\
lerman@isi.edu}

\maketitle

\begin{abstract}
As the rate of content production grows, we must make a staggering number of daily decisions about what information is worth acting on. For any flourishing online social media system, users can barely keep up with the new content shared by friends. How does the user-interface design help or hinder users' ability to find interesting content? We analyze the choices people make about which information to propagate on the social media sites Twitter and Digg. We observe regularities in behavior which can be attributed directly to cognitive limitations of humans, resulting from the different visibility policies of each site. We quantify how people divide their limited attention among competing sources of information, and we show how the user-interface design can mediate information spread.
\end{abstract}
\section{Introduction}
\label{sec:intro}

Cognition requires energy, of which we have a limited supply; any time we read a web-page or share a video, we expend the bit of energy required for that mental effort, and it has been well documented that our reserves for these actions are not very deep~\cite{Muraven:1998ta,Baumeister:2008ge,blus:2008tn,Sarter:2006dl,Smit:2004jo}.  As a result, we possess limited endurance for processing incoming information, realized as limited attention~\cite{Kahneman73,Rensink:1997vj,Pashler:1998ug}. The interaction between our limited attention and an ever-growing volume of information, especially user-generated information in online social media, has non-trivial consequences on how people consume and share information. Understanding this interaction is essential to understanding how information propagates online and how user-interface design mediates the online social experience.

So far, few computer scientists have addressed the problem of limited divided attention~\cite{Goldhaber97,Weng:2012dd,Hodas12socialcom}, especially under conditions of information overload.  Existing works, particularly~\cite{Hodas12socialcom}, showed that information in social media competes for attention of the user, and content's visibility \remove{position on the page} is vital to it being spread. Specifically, users start at the top of the web page, working their way down to content at the bottom of the page or on other pages~\cite{Counts11}. Taken alone, this pattern in human perception does not provide any insight into the problem, because, hypothetically, the user could keep browsing until all available content is consumed, expending as much effort as necessary to do so.  In practice, this rarely happens, because users exhaust their energy, loose interest, or get distracted by other tasks long before they process all available content. Nevertheless,  this super-human assumption is routinely made in social media analysis, either implicitly~\cite{Wu07} or explicitly~\cite{Kitsak10}.
%In fact, it is not \emph{a priori} given how our cognitive limitations will be realized in any scenario, necessitating the careful study of real-world examples.

In the present work, we explore how our cognitive limitations affect information spread on two social media platforms, Digg and Twitter. Users utilize social media for a variety of purposes, ranging from pure information broadcasting to pure information consumption.  By separating users into (as homogeneous as possible) populations and by aggregating behaviors over sufficient sample size, we average out message-specific and user-specific factors,  revealing behavioral traits statistically common to the population as a whole.  Because the websites for these two platforms utilize different visibility policies by default, we reveal how information competes for attention and how the position on the web-page determines uptake.  Our study reveals common patterns of human behavior.  People pay more attention to recent content, but primarily because it is easiest to find.  Controlling for visibility, we find little evidence that older content (up to one week old) is inherently less appealing to users.  Despite the differences between the two social networks, controlling for visibility shows that users rapidly reach their cognitive limits, and we should not assume the average person has the time or energy to fully utilize social networks beyond the most highly visible content.

\section{Results}
\label{sec:socmed}
Social media sites allow registered users to create content, in the form of videos, photographs, or text messages. Users can also elect to follow the activity of other users, i.e., to see the content friends created or posted recently. For example, Twitter allows users to post and read short messages, called tweets, which often contain embedded URLs to external web content. Digg, before its major redesign in 2012, specialized in news stories. It allowed users to submit URLs linking to news stories and vote for, or \emph{digg}, those stories. Both sites also allow users to designate friends and follow their activities. The friend relationship is directed: when \emph{Bob} lists \emph{Alice} as a \emph{friend}, \emph{Bob} can see the stories \emph{Alice}recommended  but not vice versa. We call \emph{Bob} the follower of \emph{Alice}. By submitting a story or posting a message, \emph{Alice} exposes her followers, including \emph{Bob}, to the message. \emph{Bob} becomes ``infected'' by voting for the story or retweeting the message, exposing his own followers to it.
%{In the event that $j$ has $n$ friends who have voted for a story, the story appears in their interface with a colored badge with the number $n$ emblazoned on it.}
Both Digg and  Twitter's interface employ a last-in-first-out queue, so the most recently posted content is at the top of the screen, with older information ordered chronologically. However, on Twitter each successive retweet of the URL reappears at the top of the follower's stream on the Twitter website, so a URL may appear multiple times in stream. In contrast, on Digg the story remains in the same order in a user's stream, ordered by the time of any friend's first vote, i.e. ordering is based only on the first appearance in a user's stream.  Although a URL will appear only once, Digg updates a badge next to the story to reflect the number of friends who have voted for the story.

\subsection{Datasets}
We used the Digg API to collect data about 3.5K stories promoted to the front page in June 2009 and the times at which 140K distinct users voted for these stories.\remove{The data set is available at\\http://www.isi.edu/\textasciitilde lerman/downloads/digg2009.html} We also collected all voters' friends, giving us a social graph with 280K users and 1.7M links. We analyze only the votes these stories received \emph{before} promotion to the front page. Before promotion, URLs are found on Digg through the social network's friend interface (by seeing the stories friends liked recently), or on the Upcoming stories pages. It is unlikely that the Upcoming stories section is an important source for discovering new content, because it received tens of thousands of new submission daily.

We used Twitter's Gardenhose API to collect tweets over a period of three weeks in the Fall of 2010 --- roughly 20\%-30\% of all user activity at the time data was collected. We retained tweets that contained a URL in the body of the message. We used Twitter's search API to retrieve all tweets containing  those URLs, ensuring the complete tweeting history of all URLs, giving us more than 3 million tweets in total. We also collected the friend and follower information for all tweeting users, resulting in a social graph with almost 700K nodes and over 36M edges. We filtered out URLs whose retweeting behavior exhibits patterns associated with spam~\cite{Ghosh2011snakdd}. This `spam-filtered' data set contained 2072 distinct URLs retweeted a total of 1337K times by 487K distinct users.

%3,798 distinct URLs retweeted by 542K distinct Twitter users.
%We analyze the tweets from over 700,000 active users\footnote{An active user sent at least 1 tweet over the course of the observation period.} over 1 month, from blah to blah.  We tracked the usage of over 60,000 unique URLs.   Each URL forms a cascade~\cite{}, and

\subsection{Exposure response}
Using collected data, we can trace information flow through the follower graphs of Twitter and Digg.  We use time stamps in the tweets (or Digg votes)  combined with the follower graph to track when people are \emph{exposed} to some information, i.e., when a specific URL appears in their stream. The timestamps also record when they become \emph{infected} by re-broadcasting it via the social network. We consider a infection to be a retweet of a URL on Twitter or a vote for it on Digg.\footnote{We define a retweet/vote by a user to be any broadcast of a URL that had previously appeared in her stream because it was tweeted/voted for by a friend previously.}
We denote user $u$ as exposed to a URL if at least one of $u$'s friends (the users followed by $u$) had previously broadcast it, regardless of whether or not $u$ consciously consumed the URL. On Twitter, multiple exposures of the same URL are treated as distinct messages, each displayed at the top of a user's stream at the time of each exposure. To eliminate the potentially confounding effects of multiple exposures, we study only those Twitter events where a user is exposed once and only once to a URL~\cite{Hodas12socialcom}, where each event is a user-exposure pair. On Digg, however, a URL's position within a user's stream is set by the time of the \emph{first} exposure by any friend. Subsequent exposures do not change its position within a user's default stream. Therefore, in Digg analysis we consider all exposures to the URL to accumulate better statistics.

%\paragraph{Infections vs Exposures}
\begin{figure}[tbp] %  figure placement: here, top, bottom, or page
 \begin{tabular}{@{}c@{}c@{}}
       \includegraphics[width=0.5\columnwidth]{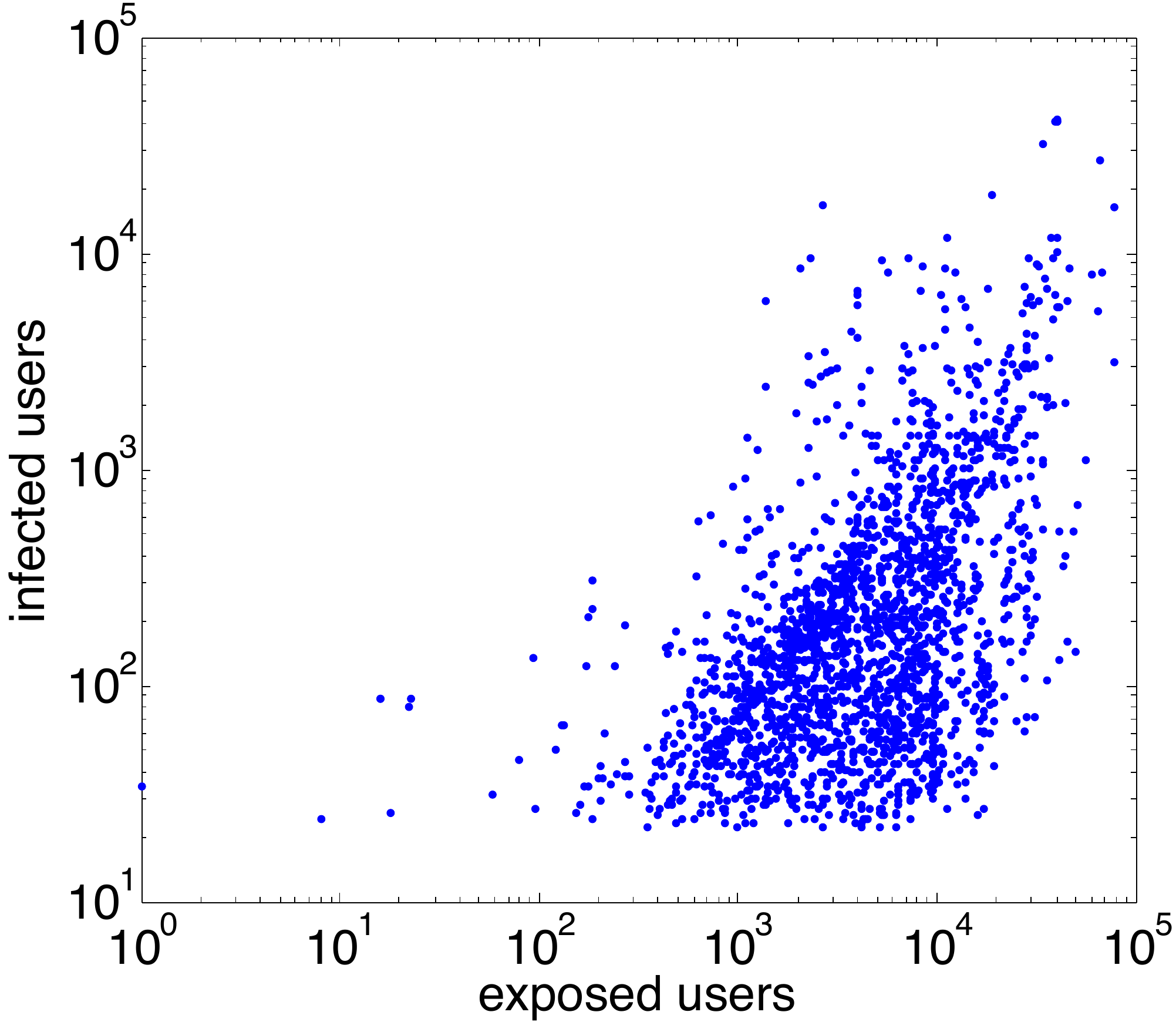} &
  	 \includegraphics[width=0.5\columnwidth]{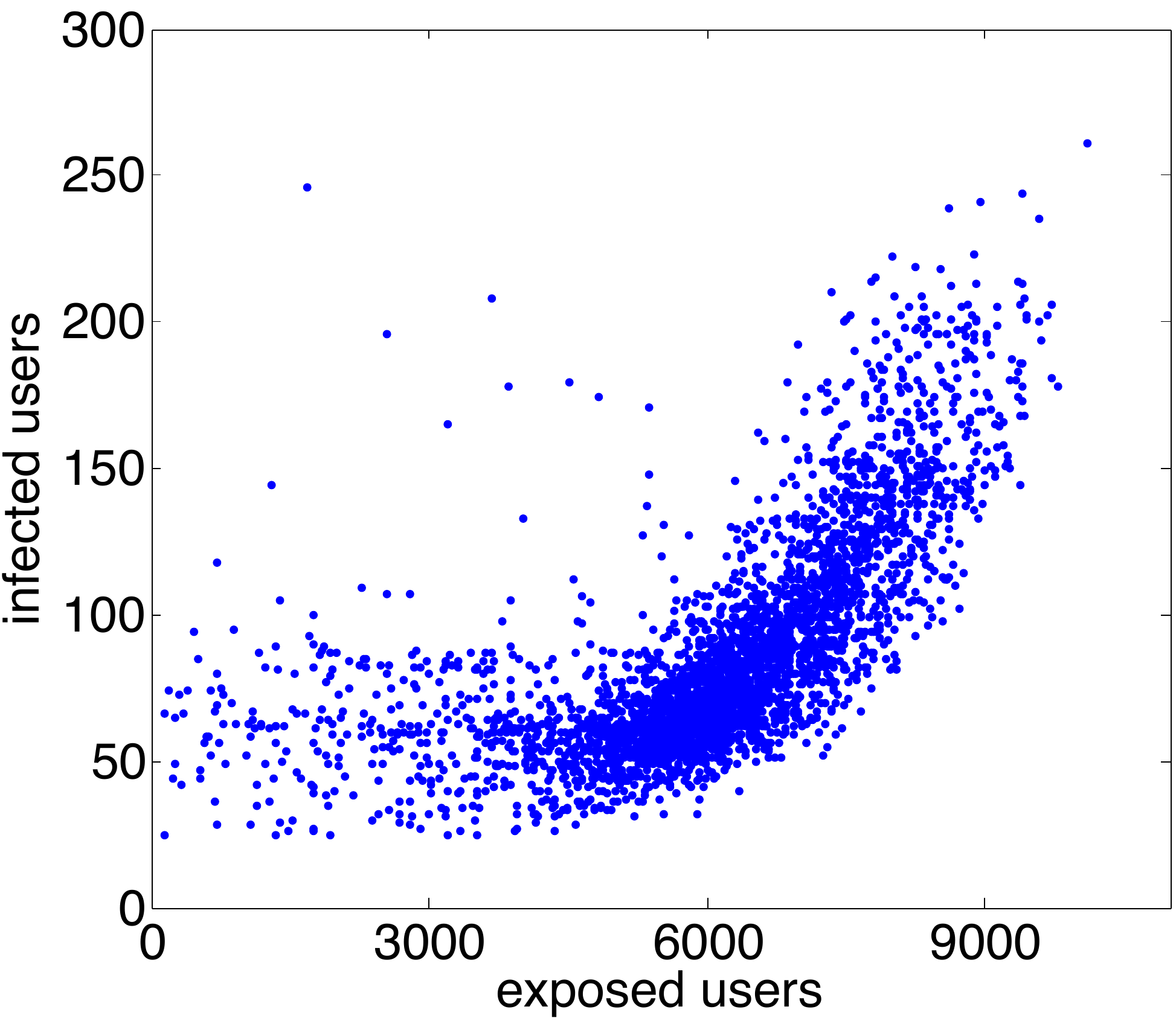}\\
	 (a) Twitter & (b) Digg
  \end{tabular}
  \caption{Number of users who are infected by a URL, because they (a) retweet it on Twitter or (b) vote for it on Digg, as a function of the number of users who are exposed to the URL on the site.}
   \label{fig:exposure}
 \end{figure}

Figure~\ref{fig:exposure} shows the number of users exposed to each URL in our dataset and the number of infected users generated by these exposures. On average, the number of infections increases super-linearly with the number of exposures, although with a large variance.   If users solely selected content to spread based on interestingness -- regardless of the difficultly of finding the content -- we would expect infections to be proportional to (or even independent of) exposures. Figure~\ref{fig:exposure} suggests that
%the visibility of the URL, which determines how easily people can see it, is a significant factor in how users respond to it, as shown in more detail below.
%the more exposure a URL has, i.e., the more easily it can be found, affects how users respond to it, as shown in more detail below.
the more easily a URL is found (more visibility), the more users respond to it. We examine this in detail below.

\begin{figure}[tb] %  figure placement: here, top, bottom, or page
 \begin{tabular}{@{}c@{}c@{}}
  	 \includegraphics[width=0.5\columnwidth]{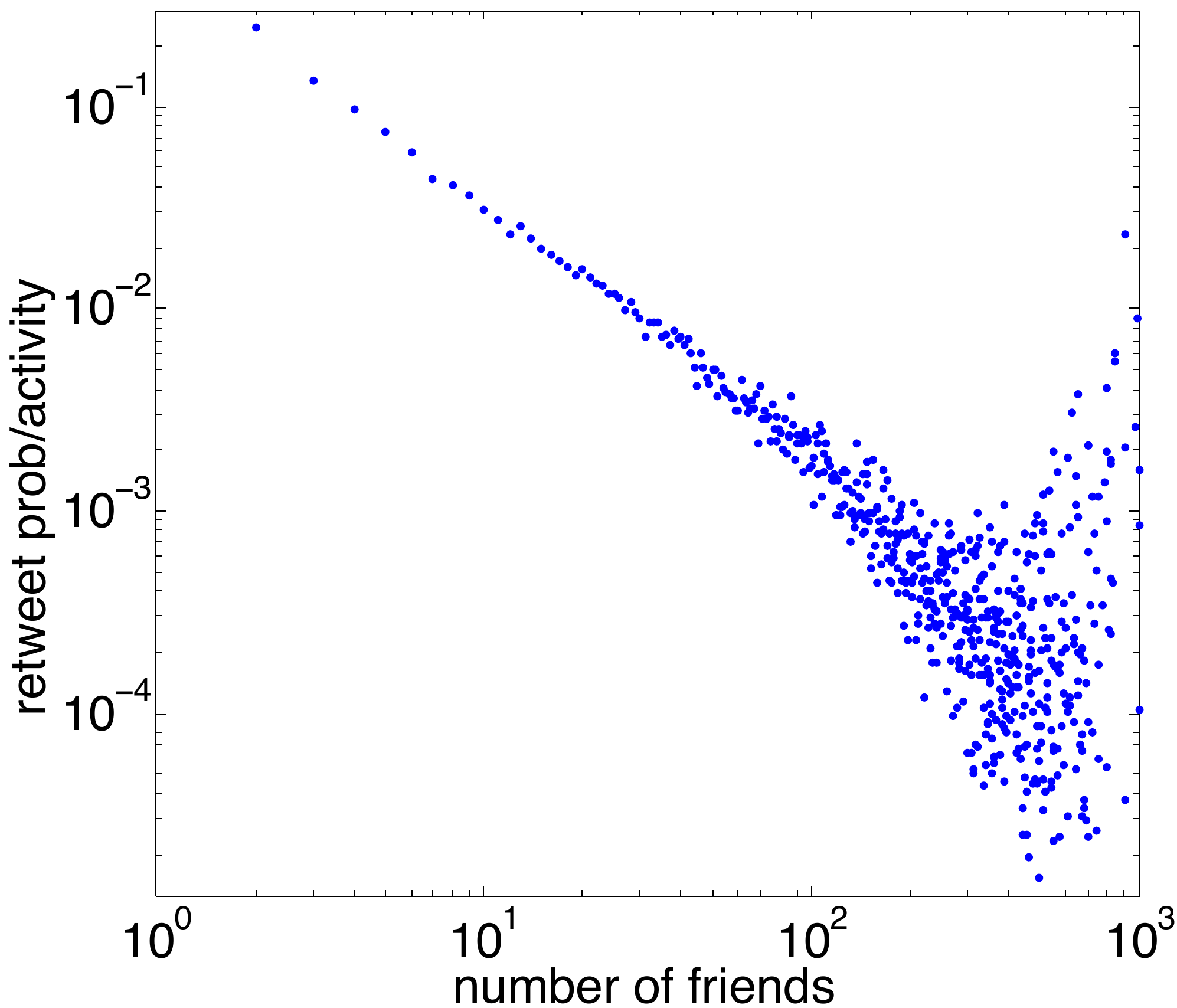} &%{./Figures/Twitter/friendnormalization_entropy} &
 	 \includegraphics[width=0.5\columnwidth]{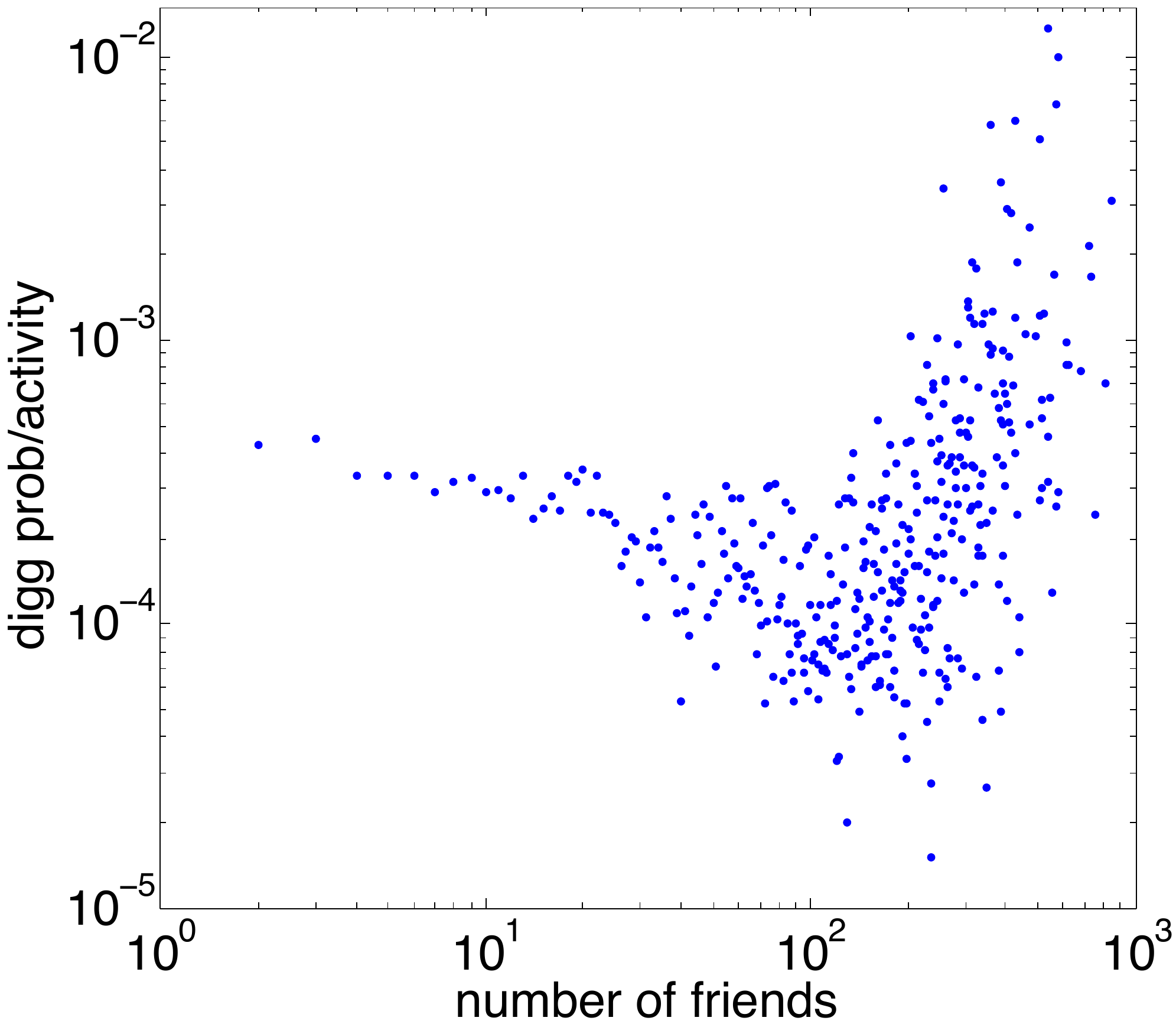}\\
	 (a) Twitter & (b) Digg
% 	  \label{fig:noveltynormalization}
\end{tabular}
   \caption{The probability of retweeting a URL given a single exposure drops off  monotonically as a function of the number of friends the user follows, which indicates that the user's attention is divided among incoming messages.
   }
 \label{fig:friendnormalization}
   \end{figure}

\subsubsection{Divided attention}
Deciding which URLs to re-broadcast is an attentive task  requiring mental effort.  Since the human brain has a limited energy budget, its ability to process new stimuli is also limited~\cite{blus:2008tn}, resulting in the phenomenon of limited attention.  Moreover, online, as in real life, people often divide their limited attention among competing incoming messages or stimuli. On social media limited divided attention reduces the likelihood that a highly-connected user will respond to a specific message from a friend~\cite{Hodas12socialcom}. Moreover, adding friends, on average, results  in a super-linear increase in incoming information and reduces a users responsiveness~\cite{HodasICWSM2013}.
%Moreover, the more friends a user has,  the less likely she is to respond.  %In \cite{Hodas12socialcom} we  demonstrated that a Twitter user's sensitivity to a single exposure to a URL is inversely proportional to the number of friends she follows.
This happens because the user must search through all messages to find a specific one that contains some URL. The more friends she has, the faster new messages arrive in the user's stream, which means she will have more difficulty finding messages which appear only once~\cite{HodasICWSM2013}.  This constraint holds even if a user wishes to give selective attention to specific friends or content.
In this paper we show that the phenomenon of divided attention also exists on Digg, and it looks very similar to Twitter.

%The effect of divided attention can already be seen in the time response function: probability of responding to a URL drops off faster in time for highly connected users than for poorly connected users.
We quantify the effect of divided attention by measuring the probability of responding to a URL  as a function of the number of friends, $n_f$, the user follows. Although we do not know all of the messages a user receives, we know that it will scale $\sim n_f^\alpha$, where $\alpha$ has been measured as 1.14~\cite{HodasICWSM2013}.  Response probability on Twitter is computed for events where users were exposed once to the URL.  For Digg it is computed for all events.  Because users with larger $n_f$ spend more time using Twitter and/or Digg on average, we normalize the probability of responding by average activity, defined as the mean number of messages sent by users  with the given $n_f$.  Figure~\ref{fig:friendnormalization} shows that this probability decays with  $n_f$ on both Twitter and Digg.   Thus, per unit activity, there is a systematic decrease in response probability to any given URL as $n_f$ increases. The upward trend in the data for users with high $n_f$ is mostly due to a sparsity of observed events, leading to a rising lower-bound on the estimable likelihood of activity,  disrupting the greater observed trend.
%The solid line in Fig.~\ref{fig:friendnormalization}(a) is the fitted function
%\begin{equation}\label{eq:friendnormalization}
%\mathcal{P}_{n_f} = 0.22\,{n_f^{0.21}}/{(n_f+0.53)}.
%\end{equation}
%The factor of $n_f^{0.21}$ produces a small correction accounting for the higher Twitter activity in users with more friends.  The friend-dependent  response probability  is dominated by the inverse of the number of friends in both cases.

%most likely due to a lack of data

%Although, on Digg, for $n_f>100$, voting probability goes up again, probably due to top users. This is the elite group of top-ranked users who are abnormally active and follow many friends.

\subsection{Time response}
The probability to respond to, or be infected by, an exposure decays quickly with time.
We define the probability of an infection after a time interval $\Delta t=t-t_0$ following an exposure: $P_t(\Delta t) = \langle \mathbbm{1}_{u_i,t_0 + \Delta t} \mathbbm{1}_{u_i^\prime,t_0}\rangle_{u,t_0}$. Here $\mathbbm{1}_{u_i^\prime,t_0}=1$ if user $u_i$ is exposed, i.e., receives a URL, at time $t_0$ and 0 otherwise, and given that $u_i$ follows $u_i^\prime$ and $t>t_0$.
To calculate this function for Twitter, we first select all infections that occur after receiving a URL only once.  For Digg, we consider all exposures.  Next, we calculate a normalized histogram of the set of responses with delays $\Delta t$, given that an exposure occurred at time $t_0$ \emph{and} given that the user did respond. We denote this conditioned probability distribution  as $\mathfrak{T}(\Delta t|\chi)$, where $\chi$ is the condition satisfied by the population of users under consideration.  In general, there is less than 100\% probability that a user will respond to a given message.  The proper normalization is found by calculating the probability that a user with characteristic $\chi$ eventually responds, given that the user was exposed to that URL,
\begin{align}\label{eq:timenormalization}
\mathcal{P}_{n_f}& = \int_0^\infty P_t(\Delta t|\chi)\,d\Delta t  = \\\nonumber
& \frac{\sum_{u,j}\mathbbm{1}_{u\in\chi} \mathbbm{1}_{|V^{+,tw}(u,j)| = 1}}{\sum_{u,j}\mathbbm{1}_{u\in\chi}\mathbbm{1}_{|V^{+,tw}(u,j)| = 1}+\mathbbm{1}_{u\in\chi}\mathbbm{1}_{u\in\mathcal{W}}\mathbbm{1}_{|V^{-,tw}(u,j)| = 1}}.
\end{align}
% friend normalization
We calculated this normalization as a function of number of friends, which is a fit to data shown in Figure~\ref{fig:friendnormalization}.

\begin{figure}[tbp] %  figure placement: here, top, bottom, or page
 \begin{tabular}{@{}c@{}c@{}}
  	 \includegraphics[width=0.5\columnwidth]{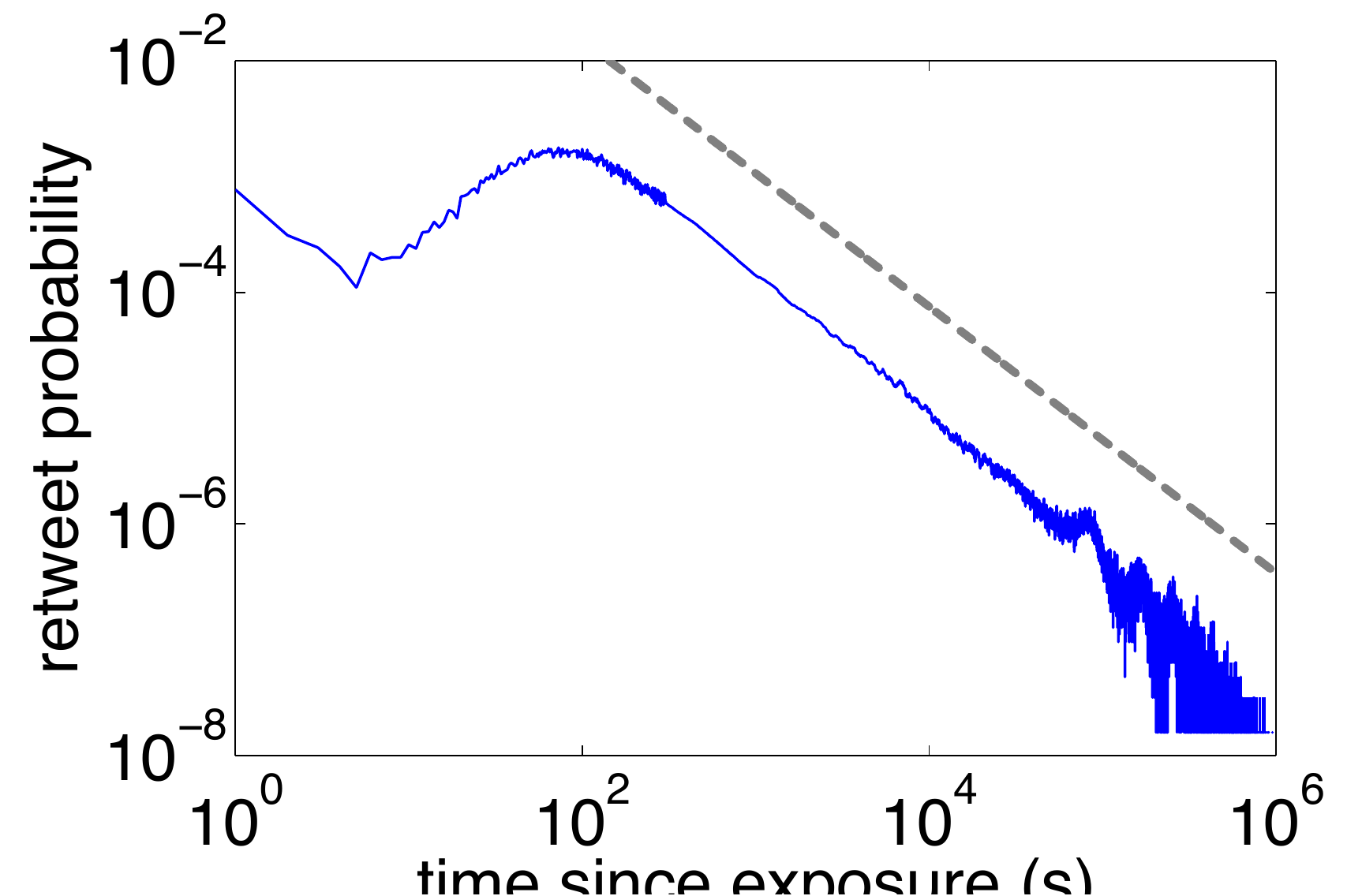} &
  	 \includegraphics[width=0.5\columnwidth]{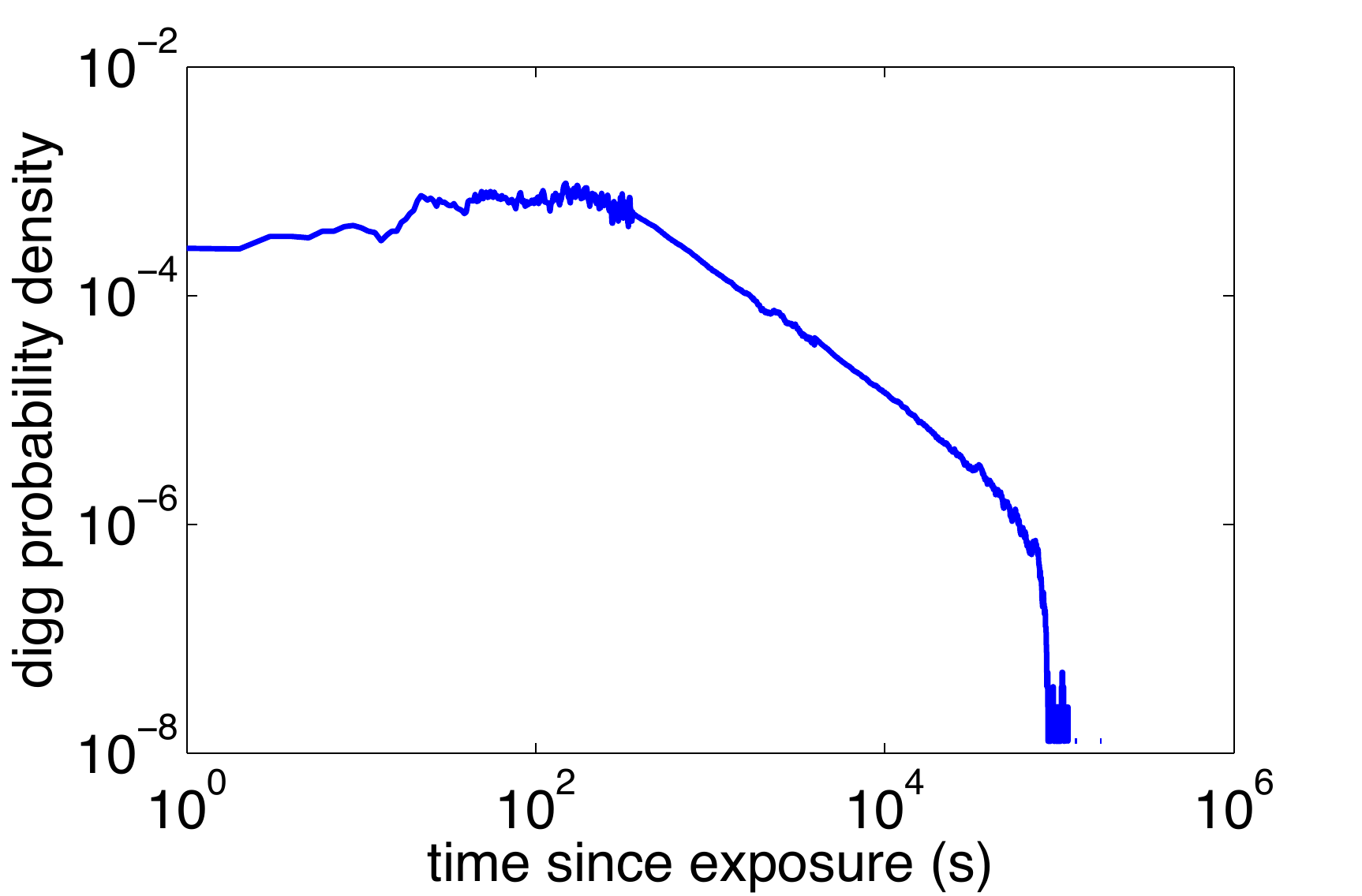}\\
	 (a) Twitter & (b) Digg
  \end{tabular}
  \caption{Time response function for users who respond to the URL after being exposed to it by a friend on (a) Twitter and (b) Digg. \remove{Black line shows average response of all users, blue line shows response of poorly-connected users, and red line shows highly-connected users.} The dashed line in (a) is $\propto\Delta t^{-1.15}$.
}
   \label{fig:timeresponse}
 \end{figure}

Figure~\ref{fig:timeresponse} shows the  time response function  $\mathfrak{T}(\Delta t | \chi)$ for different populations $\chi$ of users on Digg and Twitter. The populations are poorly-connected users with fewer than 10 friends (blue line), highly-connected users with hundreds of friends (red line), and for Digg, medium connected users (green line).  All probabilities have unit normalization and are smoothed as follows: Between 0 and 300 seconds is raw data; between 301 and 33,000 seconds is a 300 sec. moving average; and from 33,001 onward is a 3000 sec. moving average.  

On both sites,  users usually retweet or vote for the URL mere minutes after a friend has added it to their stream. After this time, infection probability drops off significantly.  We observe a broad peak \remove{on Twitter} around 2 minutes, which likely corresponds to users retweeting/digging the URL after reading the referent Web page.   After this characteristic reading time, the Twitter time response function drops off roughly as $\Delta t^{-1.15}$, shown for comparison as a dashed line. The power-law exponent in the tails of different populations of users are nearly identical, indicating that the same behavioral strategies are at play for all classes of users~\cite{Hodas12socialcom}.  Because we only analyze pre-promotion dynamics on Digg, users only have up to 24 hours to digg a URL, forcing a sharp cutoff in the time-response function.
%Differences in the time decay of different user populations on Digg indicate their behavior differences; highly connected users are much more active and highly utilize the friend interface, but they more likely to be swamped by stories.

\subsubsection{Visibility}
We interpret the time response function in terms of decay of the `\emph{visibility}' of a URL. A user is most likely to be infected by a URL  when it is easiest to find: soon after a friend has broadcast it, when it is still near the top of a user's screen.
As soon as this URL is added to a user's stream, new messages from friends push it down.  Because users pay more attention to content near the top of the screen~\cite{Buscher09}, messages that are farther from the top are more difficult to find, reducing the likelihood that a user will see them and respond to them.
\remove{This effect is more dramatic for highly-connected users (dashed red lines in Fig.~\ref{fig:timeresponse}), whose probability to respond decreases faster in time than for users with few friends.} This is consistent with the phenomenon of divided attention. Also, given that a highly connected user does respond to a message, she responds faster than a poorly connected user~\cite{Hodas12socialcom}. If she is too slow to come to the site, she may not see the message and hence not respond to it.

\subsubsection{Novelty}

\begin{figure}[htb] %  figure placement: here, top, bottom, or page
% \begin{tabular}{@{}c@{}c@{}}
%  	 \includegraphics[width=0.5\columnwidth]{./Figures/Twitter/timeactivity_entropy.jpg} &
% 	 \includegraphics[width=0.5\columnwidth]{./Figures/Digg/article_age_probability.pdf}\\
%	 (a) Twitter & (b) Digg
%\end{tabular}
 \begin{tabular}{@{}c@{}}
  	 \includegraphics[width=\columnwidth]{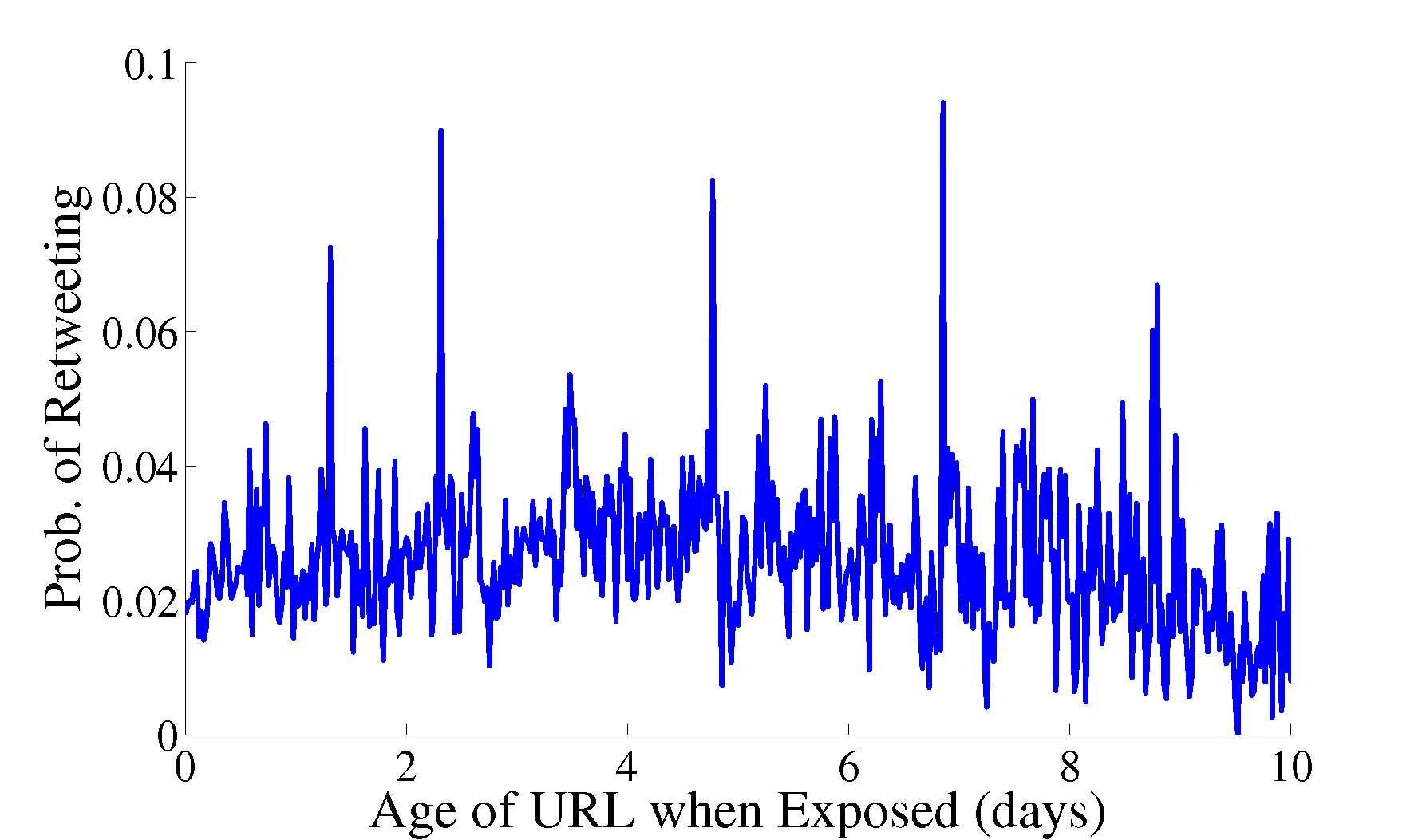} \\
	 (a) Twitter \\
 	 \includegraphics[width=\columnwidth]{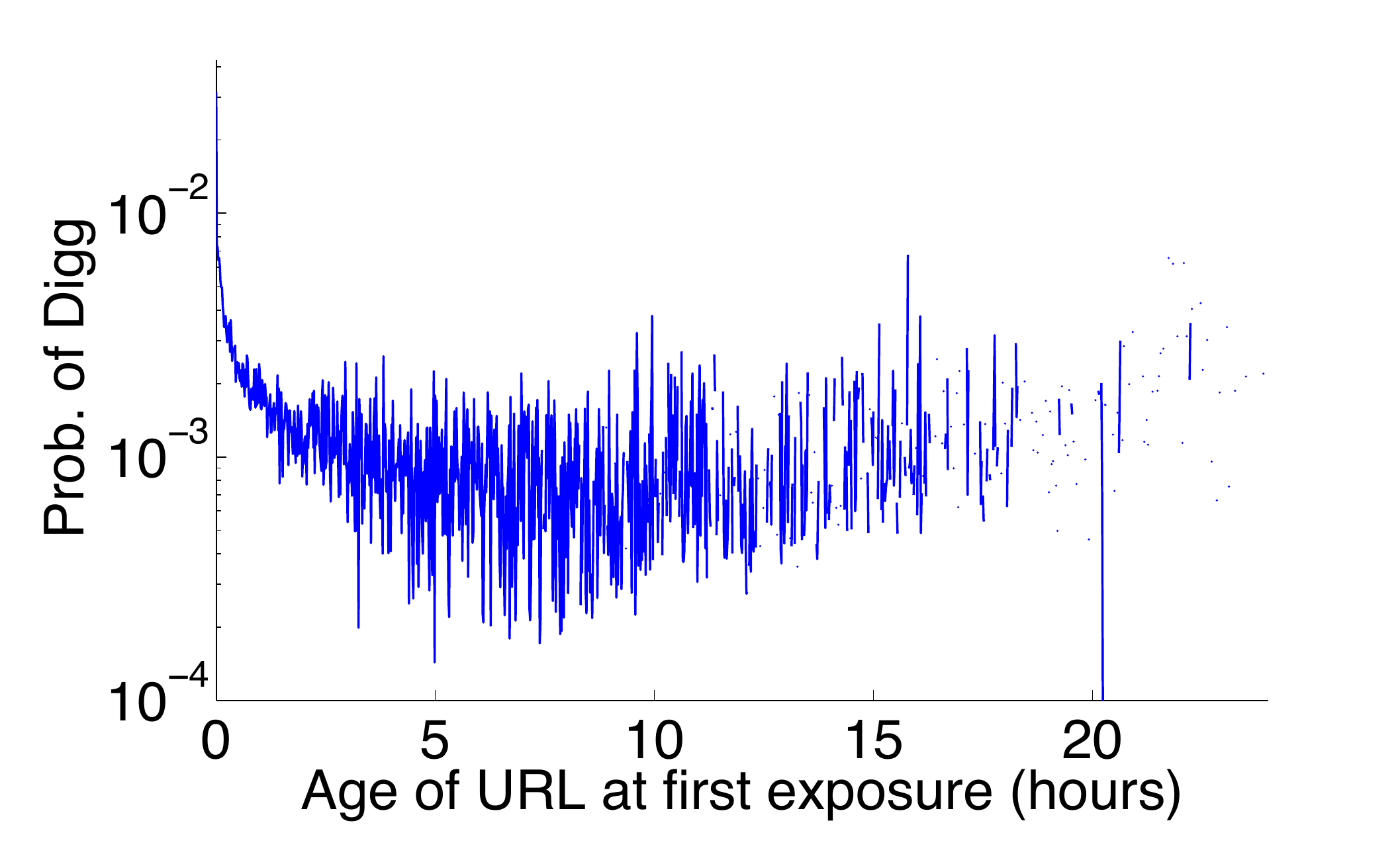}\\
	  (b) Digg
\end{tabular}
   \caption{The probability of responding to a URL as a function its age (a) in days on Twitter (with 30 minute binning) and (b) for the period of 24 hours on Digg.}
 \label{fig:noveltydecay}
   \end{figure}

In the past, researchers have interpreted the time-dependent decrease in infection probability as decay of the novelty of content~\cite{kleinberg09news,Moussaid09,Wu07}. However, temporally decaying visibility will also lead to temporally decreasing infection probability. In this section we examine whether novelty bias exists in social media. In other words, do people pay more attention to content because it is more novel or because it is more visible?

To answer this question, we calculate the infection probability for URLs of a given age $\tau$. This is formalized as:
\begin{align}\label{eq:timenormalizationNovelty}
\int_0^\infty &P_t(\Delta t|\tau)\,d\Delta t  = \\ \nonumber
&\frac{\sum_{u,j} \mathbbm{1}_{|V^{+,tw}(u,j|\tau)| = 1}}{\sum_{u,j}\mathbbm{1}_{|V^{+,tw}(u,j|\tau)| = 1}+\mathbbm{1}_{u\in\mathcal{W}}\mathbbm{1}_{|V^{-,tw}(u,j|\tau)| = 1}},
\end{align}
\noindent where $\tau$ is the time since the first appearance of the URL in the dataset. $|V^{+,tw}(u,j|\tau)|$ is the number of messages received by a user $u$ containing URL$_j$, given that the last tweet (first vote) received by $u$ containing URL$_j$ was at $\tau$ and $u$ did rebroadcast URL$_j$. Equivalently, $|V^{-,tw}(u,j|\tau)|$ is the number of messages received by a user $u$ containing URL$_j$, given that the last tweet (first vote) received was at $\tau$ but $u$ did not rebroadcast URL$_j$.

Figure~\ref{fig:noveltydecay} shows the calculated probabilities as a function of URL age on Twitter and Digg. For the Twitter data, we calculated these probabilities for URLs ranging in age from one to ten days~\cite{Hodas12socialcom}. As we can see from the figure, the absolute age of the URL when seen by a user has little effect on its retweet probability. There also appears to be no novelty decay on Digg, though the evidence for it is less clear. We are only monitoring response to a URL before it is promoted to the front page, which happens within 24 hours of its first appearance in the dataset, which is too short a period to validate novelty decay. We do not attempt to ascertain novelty decay for URLs after they are promoted, because front page significantly boosts story visibility. Ref.~\cite{Hogg12epj} explains how the popularity of stories dynamically depends on their visibility on the front page.

\section{Related Work}\label{sec:relatedwork}

The relation of human attention to individual and consumer choice has been well studied over many years, although generally in the context of controlled laboratory experiments~\cite{Kahneman73,Anderson09,blus:2008tn,Falkinger2007Attention}. Attention has also been invoked to explain online social behavior~\cite{Goldhaber97}.  For example, the collective shifts in popularity between events and topics over time has been referred to as ``collective attention"~\cite{Wu07,Wilkinson08,Huberman-attention,Moussaid09,Ratkiewicz10,Weng12}.  Previous efforts have been made to better understand how individuals utilize their perceptive abilities to process incoming information, such as~\cite{Buscher09,Counts11}, which showed that users concentrate on content near the top of the screen. Fewer studies have addressed divided attention, the phenomenon that as the number of information sources grow, people allocate less attention to each source.  A study of conversations between Twitter users found that people limit themselves to 150 or so conversation partners~\cite{Goncalves11}. Twitter users were shown to divide their attention among all incoming messages, regardless of the content or quality of the underlying messages~\cite{Hodas12socialcom}.

\section{Conclusion}\label{sec:conclusion}

As inexpensive personal phones have evolved into smart multimedia devices, people all over the world have been given the tool to generate and share media on an unprecedented scale --- a scale which will only continue to rapidly expand.  If users are not at information overload, they soon will be.  How should information be displayed to users to minimize information overload yet maximize the potential to uncover novel content?  The answer demands a deeper understanding of individual perception tendencies and divided attention, which is the core focus of the present work.   By comparing two popular social media sites, we demonstrated how to analyze statistical user behavior to uncover the effects of the visibility policy of the site itself.

Our study identifies common patterns of human behavior.  Some conclusions may not be surprising, but only in hindsight, given the abundance of models attempting to describe information contagion.  Our quantified results eliminate some previous assumptions, such as the relevance of ``novelty decay" for understanding average user behavior.  A common assumption ---  users are `rational' agents fully balancing incoming information to form an optimized decision --- fails to explain the difference in spreading behavior between Digg and Twitter.

Although the present results primarily consider the impact of receiving a single URL, bypassing any possible confounding factors due to complex contagion, we can conclude that visibility and divided attention play a prominent role in determining information propagation.  Further work will elucidate how to better quantify the relative importance of screen position and how saliency can be manipulated to aide users in uncovering valuable information.

-------------------
\bibliographystyle{IEEEbib}
\bibliography{biblibrary_standard}

\end{document}